\documentclass[pra, twocolumn, showpacs]{revtex4}
\usepackage{amsmath}
\usepackage{amssymb}
\usepackage{graphicx}
\usepackage{stackrel}
\usepackage{color}
\usepackage{subfigure}

\newcommand{\br}{{\bf r}}
\newcommand{\bR}{{\bf R}}
\newcommand{\be}{\begin{equation}}
\newcommand{\bea}{\begin{eqnarray}}
\newcommand{\ee}{\end{equation}}
\newcommand{\eea}{\end{eqnarray}}
\newcommand{\om}{\omega}
\newcommand{\bk}{{\bf k}}
\newcommand{\bE}{{\bf E}}
\newcommand{\bF}{{\bf F}}
\newcommand{\bp}{{\bf p}}
\newcommand{\bff}{{\bf f}}
\newcommand{\beo}{{\hat{\bf e}_0}}
\newcommand{\ber}{{\hat{\bf r}}}
\begin{document}

\title{Consistency of a Causal Theory of Radiative Reaction\\ 
with the Optical Theorem}

\author{F. Intravaia and R. Behunin}
\affiliation{Theoretical Division, Los Alamos National Laboratory, Los Alamos, NM 87545, USA}
\author{P. W. Milonni}
\affiliation{Theoretical Division, Los Alamos National Laboratory, Los Alamos, NM 87545, USA}
\affiliation{Department of Physics and Astronomy, University of Rochester, Rochester, NY 14627, USA}
\author{G. W. Ford}
\affiliation{Department of Physics, University of Michigan, Ann Arbor, MI 48109-1120, USA}
\author{R. F. O'Connell}
\affiliation{Department of Physics and Astronomy, Louisiana State University, Baton Rouge, LA 70803-4001, USA}

\begin{abstract}
The Abraham-Lorentz-Dirac equation for a point electron, while suffering from runaway solutions and an acausal response to external forces, is compatible with the optical theorem. We show that a theory of radiative reaction that allows for a finite charge distribution is not only causal and free of runaway solutions, but is also consistent with the optical theorem and the standard formula for the Rayleigh scattering cross section.
\end{abstract}

\pacs{42.50.Nn --Quantum optical phenomena in absorbing, amplifying, dispersive and conducting media; cooperative phenomena in quantum optical systems, 05.40.-a --Fluctuation phenomena, random processes, noise, and Brownian motion, 42.50.Lc --Quantum fluctuations, quantum noise, and quantum jumps}
\maketitle

The theory of radiative reaction leading to the Abraham-Lorentz-Dirac (ALD) equation, while exhibiting such notorious features as runaway solutions and preacceleration \cite{LandauJackson}, is nevertheless consistent with the optical theorem and the Rayleigh scattering cross section. One approach to the resolution of the problems besetting that theory is based on the quantum Langevin equation describing dissipative quantum systems \cite{Fordkac}, together with the assumption that the electron is not a point particle but is described by a form factor with a very high cutoff frequency \cite{Ford91,rfo}. The classical, nonrelativistic equation of motion for an electron in this theory is free of preacceleration and runaway difficulties  \cite{Jacksonagain}, and we show in this Brief Report that it is also consistent with the optical theorem and the Rayleigh cross section. 

We first recall some basic aspects of the linear response of an electron, described as a rigid charge distribution $\rho(\br)$ centered at ${\bf R}(t)$, to an applied, sufficiently small electric field
\be
{\bf E}({\bf r},t)={\bf E}_0e^{i({\bf k}_0\cdot{\bf r}-\omega t)} \ \ \ (|\bk_0|=k=\om/c).
\ee
The force exerted on the electron by this field is
\\
\bea 
{\bf F}(t)&=&\int d^3r\rho(\br-\bR(t))\bE(\br,t)\nonumber\\
&=&\int d^3r\rho(\br)\bE(\br+\bR(t),t)\nonumber \\
&\cong& \int d^3r\rho(\br)e^{i\bk_0\cdot\br}\bE_0e^{-i\om t}\nonumber \\
&=&ef(\bk_{0})\bE_0e^{-i\om t},
\eea
where $e$ and $f(\bk_{0})$ are the electron charge and form factor, respectively. We have made the dipole approximation $\bk_0\cdot\bR(t)\ll 1$: this is equivalent to saying that the size of the dipole, associated with the electron displacement, is small with respect to the wavelength of the incident radiation. This, however, does not impose any limitation on the size of the charge distribution ($f(\bk)$ is still generic).

In terms of the Fourier transforms $\tilde{\bR}(\om)$ and $\tilde{{\bf F}}(\om)$ of the electron displacement and the applied force, respectively, the linear response of the electron to the applied field is expressed as
\be
\tilde{\bR}(\om)=\alpha(\om)\tilde{\bF}(\om)=ef(\bk_{0})\alpha(\om)\bE_0.
\label{Req}
\ee
The function $\alpha(\om)$ is determined by the equation of motion for the electron in the presence of the applied field and
any additional forces acting upon it. Its
calculation with radiative reaction can be far from trivial \cite{pwm1}, but it is
well known that, with or without radiative reaction, it must satisfy certain basic conditions:
\begin{equation}
{\alpha}(\omega)={\alpha}^{*}(-\omega),
\label{crossing}
\end{equation}
and, as a function of complex frequency $\zeta$,
\begin{equation}
{\alpha}(\zeta)\text{ is analytic  for } \mathrm{Im}\,  \zeta>0.
\label{analyticity}
\end{equation} 
The first condition, the ``crossing relation," is simply the requirement that the induced dipole moment is real. The second is a direct consequence of causality and implies the familiar Kramers-Kronig relations between the real and imaginary parts of the polarizability \cite{LandauJackson}.

Scattering theory provides a further constraint in terms of the \emph{optical theorem} \cite{Jackson2} relating the total scattering cross section and the forward scattering amplitude: 
\begin{equation}
\sigma_t=\frac{4\pi}{k}{\rm Im}[\beo^*\cdot\bff(\bk_0,\bk_0)].
\label{opttheorem1}
\end{equation}
This is just the requirement of energy conservation, or, in quantum theory, the conservation of probability. 
In our case, to calculate  $\bff(\bk_0,\bk_0)$ it is sufficient to consider the electric field emitted by the electron in the radiation zone \cite{Jackson2}: for a confined 
current density ${\bf j}(\br,\om)$ the field is
\begin{subequations}
\begin{align}
&\bE(\br,\om)=k^2[\bp(\mathbf{k}_{s},\om)-(\ber\cdot\bp(\mathbf{k}_{s},\om))\ber]\frac{e^{ikr}}{r},\\
&\bp(\mathbf{k}_{s},\om)=\frac{i}{\om}\int{\bf j}(\br',\om)e^{-\imath\mathbf{k}_{s}\cdot\br'}d^3r' =\frac{i}{\om}\tilde{\bf j}(-\bk_{s},\om),
\end{align}
\end{subequations}
where $\tilde{\bf j}(\bk,\om)$ is the space-time Fourier transform of the current distribution and $\bk_{s}$ is the wavevector in the direction of observation ($\bk_{s}=k\ber=k\br/r$). Writing
\be
{\bf j}(\br,t)=\dot{\bR}(t)\rho(\br-\bR(t)),
\ee
and again making the dipole approximation $\bk\cdot\bR(t)\ll 1$, we have \cite{Ford91a}
\be
\tilde{\bf j}(-\bk_{s},\om)=-i\om e f(-\bk_{s})\tilde{\bR}(\om)=-i\om e f^{*}(\bk_{s})\tilde{\bR}(\om)
\ee
and
\be
\bE(\br,\om)=e f^{*}(\bk_{s})k^2[\tilde{\bR}(\om)-\ber\cdot\tilde{\bR}(\om)\ber]\frac{e^{ikr}}{r}.
\ee
With $\tilde{\bR}(\om)$ given by (\ref{Req}), we identify the scattering amplitude 
\be
\bff(\bk_{s},\bk_0)=f^{*}(\bk_{s})f(\bk_{0})e^2\alpha(\om)(k^2\beo-\bk_{s}\cdot\beo\bk_{s}),
\ee
where $\beo=\bE_0/E_0=\bR/R$ is the unit polarization vector of the incident field. The forward scattering amplitude ($\bk_{s}$ orthogonal to $\beo$) is therefore
\be
\bff(\bk_0,\bk_0)=k^2|f(\bk_{0})|^{2}e^2\alpha(\om)\beo,
\label{fscatt1}
\ee
The total scattering cross section $\sigma_t$ can be  obtained by integration over all solid angles is
\be
\sigma_t=\int d\Omega|\bff(\bk,\bk_0)|^2=\frac{8\pi}{3}k^4|f(\bk_{0})|e^4|\alpha(\om)|^2.
\label{fscatt2}
\ee
Hence, from (\ref{opttheorem1}) we have 
\be
{\rm Im}[\alpha(\om)]=\frac{2e^2\om^3}{3c^3}|\alpha(\om)|^2|f(\bk_{0})|^{2},
\label{optrayl}
\ee
which is an equivalent statement of the optical theorem~\cite{pwm1}. 

In the ALD theory of radiative reaction \cite{LandauJackson} a dipole oscillator subject to a restoring force $-M\om_0^2\bR$ and an external electric field ${\bf E}$ is described nonrelativistically by the equation of motion
\begin{equation}
\ddot{\bf R}+\omega_0^2{\bf R}=\frac{e}{M}[{\bf E}+{\bf E}_{RR}],
\label{al1}
\end{equation}
where $M$ and $\omega_0$ are respectively the (observed) electron mass and the resonance frequency.  
The term ${\bf E}_{RR}=(2e/3c^3)\stackrel{...}{\bf R}$ is the radiative reaction field. This implies 
\begin{equation}
\alpha(\omega)=\frac{1}{M}\frac{1}{\omega_0^2-\omega^2-i\omega^3\tau_e},\quad\text{(ALD)}
\label{alpha1}
\end{equation}
where $\tau_e=2e/3Mc^3$ is on the order of the time for light to
travel a distance equal to the classical electron radius, $r_{0}=e^{2}/Mc^{2}$. 
This expression for $\alpha(\om)$ obviously satisfies the crossing relation (\ref{crossing}), and it is also seen from
(\ref{optrayl}) that the optical theorem is satisfied with $|f(\bk)|=1$, which is the form factor for a point-like electron.  
It also follows from (\ref{fscatt2}) that, with $|f(\bk)|=1$, we have
\be
\sigma_t=\frac{[n^2(\om)-1]^2}{6\pi N^2}\left(\frac{\omega}{c}\right)^{4},
\label{raycross}
\ee
recovering the familiar, experimentally measured, Rayleigh scattering cross section for a dilute gas of isotropic, point-like scatterers: the gas refractive
index $n(\om)$ is given in this case by $n^2(\om)=1+4\pi Ne\alpha(\om)$, where $N$ is the particle number density \cite{ubachs}.

However, the result (\ref{alpha1}) of the ALD theory violates
the causality requirement that $\alpha(\om)$ be analytic in the upper half of the complex frequency plane. Additionally, as is well known, the equation of motion (\ref{al1}), from which (\ref{alpha1}) follows, exhibits runaway solutions as a consequence of the ``non-Newtonian" dependence of ${\bf E}_{RR}$ on the third derivative of ${\bf R}$.

The alternative approach to radiative reaction cited earlier \cite{Ford91,rfo}  (FO) is based on the quantum theory of dissipation in which a particle is coupled to a ``bath"  of harmonic oscillators, so that it experiences a Langevin force together with a dissipative force due to the back reaction of the oscillators on the particle. In the case of interest here the bath oscillators are associated in the usual way with the electromagnetic field, the Langevin force is due to the fluctuating electric field, and the back reaction results in the radiative damping force.  The semiclassical equation of motion for an electric dipole oscillator with bare mass $m$ is \cite{Ford91,rfo} 
\begin{equation}
m\ddot{\bf R}(t)+\int_{-\infty}^tdt'\mu(t-t')\dot{\bf R}(t')+K\mathbf{R}={\bf F}(t),
\label{fo2}
\end{equation}
which follows by taking an expectation value, so that the Langevin force, having zero expectation value, does not 
appear. ${\bf F}(t)$ is the expectation value of the externally applied force; the linearity of the system implies 
that (\ref{fo2}) describes the classical system.  
The constant $K$ characterizes a harmonic restoring force, while the function
\begin{equation}
\tilde{\mu}(\omega)=\int_{0}^{\infty}\mu(t) e^{i \omega t} \mathrm{d}t
\end{equation}
is a \emph{positive-real function}  \cite{Ford88a} and can be calculated exactly once the form factor of the electron  is provided 
\cite{Ford85,Ford87a,Ford88,Ford88a,Ford1987}. Although the exact form of $f(\bk)$ is not known, on physical grounds it can be assumed that it is unity up to some large cutoff frequency ($\Omega$), after which it falls rapidly to zero. A possible choice is \cite{Ford91,Ford85,Ford1987} 
\begin{equation}
\label{formfactor}
|f(\bk)|^{2}=\frac{\Omega^{2}}{\Omega^{2}+c^{2}k^{2}},
\end{equation}
for which \cite{Ford91, Ford1987, note1}
\begin{equation}
\tilde{\mu}(\omega)=   \frac{2e^2}{3}\frac{3c^{2} \Omega^{2}\omega}{\omega+i\Omega}, 
\label{mu}
\end{equation}
which gives the function $\mu(t)$  \cite{Ford91,rfo}:
\begin{equation}
\mu(t)=M\Omega^2\tau_e[2\delta(t)-\Omega e^{-\Omega t}],
\label{fo3}
\end{equation}
where the delta function represents the memory-less Markovian part and the second term in brackets results in non-Markovian effects. $M$ is again the observed mass of the particle and is defined here by
\begin{equation}
M=m+\frac{2e^{2} \Omega}{3c^{3}}, \text{ or } m=M (1-\tau_{e} \Omega).
\label{fo5}
\end{equation}

Various authors have connected the existence of runaway solutions of the ALD equation with a negative bare mass and the point-electron assumption \cite{erber,Ford91,Jackson2}. This is the case when $\Omega > \tau_{e}^{-1}$ (a point-like electron is recovered in the $\Omega\to \infty$ limit) for which the total energy for the system is not bounded from below. 
Requiring both the bare and renormalized masses to be positive leads to the condition $\Omega \le \tau_{e}^{-1}$. Therefore, in the large-cutoff limit, we take $\Omega= \tau_{e}^{-1}$. 
When the external force is due to an external electric field ${\bf E}({\bf r},t)$, Eqs. (\ref{fo2}), (\ref{formfactor}), and (\ref{fo3}) lead to \cite{Ford91,Ford1987}
\begin{equation}
\alpha(\omega)=\frac{1}{M}\frac{1-i\omega\tau_e}{\omega_0^2-\omega^2-i\gamma\omega}, \quad\text{(FO)}
\label{fo9}
\end{equation}
in the large-cutoff limit, where $\om_0^2=K/M$ and we have defined $\gamma=\omega_0^2\tau_e$ \cite{Ford1987}. This expression obviously satisfies the crossing relation as
well as the requirement from causality that it be analytic in the upper half of the complex frequency plane. 
From (\ref{fo9}) it also follows that
\begin{equation}
{\rm Im}[\alpha(\omega)]=\frac{2e^2\omega^3}{3c^3}|\alpha(\omega)|^2\frac{1}{1+\omega^2\tau_e^2}.
\end{equation}
Since $\Omega= \tau_{e}^{-1}$, the last factor on the right-hand side is $|f(\bk_{0})|^{2}$. Therefore the optical theorem in the form (\ref{optrayl}) for Rayleigh scattering is satisfied identically. 

The difference between equation (\ref{fo9}) and the result (\ref{alpha1}) of the ALD theory leads to
different predictions for the Rayleigh cross section (\ref{raycross}). Figure 1 compares the real and imaginary parts of these two expressions for $\alpha(\om)$ for a particular value of $\om_0$. It can be seen that, because $\tau_{e}$ is so small, both polarizabilities lead to essentially the same predictions, significant differences appearing only at extremely high frequencies at which the relativistic effects appear.
\begin{figure}[h]
\begin{center}
\subfigure[]{\includegraphics[width=3.07in]{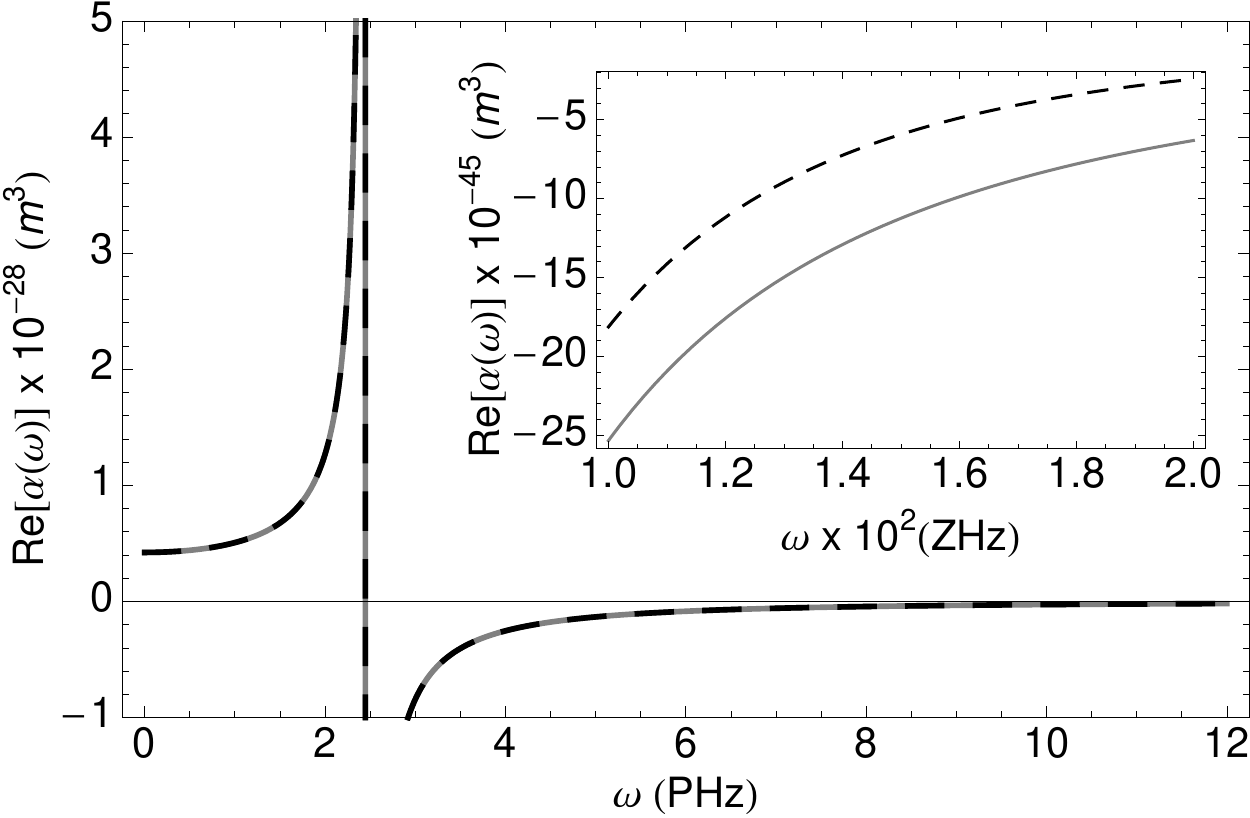}}
\quad
\subfigure[]{\includegraphics[width=3in]{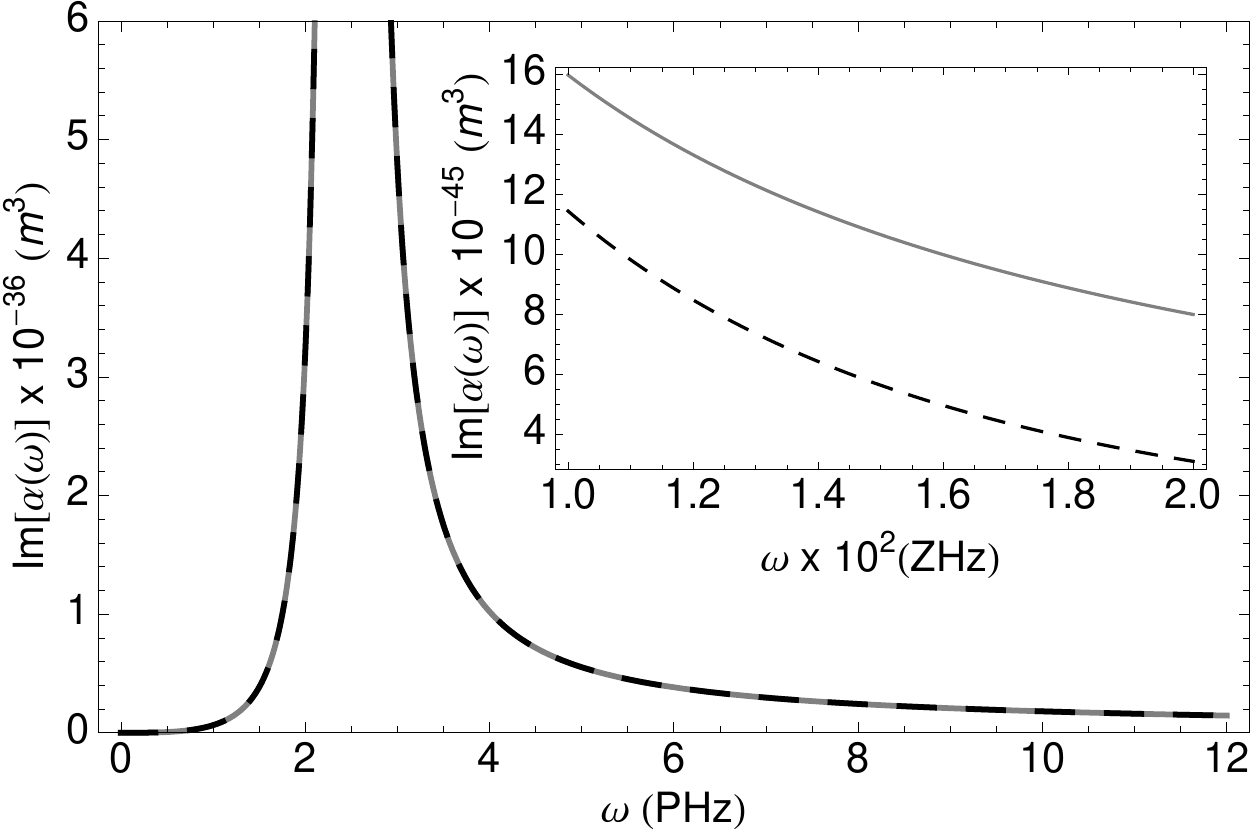}}
\caption{Real (a) and imaginary (b) parts of the polarizabilities $e\alpha(\om)$ implied by equations (\ref{fo9}) (solid curves) and (\ref{alpha1}) (dashed curves). The electron charge and mass are assumed and the oscillator's resonant frequency is set to hydrogen's first optical resonance $\omega_0 \approx 2.45 \times 10^{15}$ Hz. The insets show that significant differences between the two theories appear only at very large frequencies.}
\label{ }
\end{center}
\end{figure}

More generally, without specifying the form of $f(\bk)$, the Fourier transform of (\ref{fo2})
gives the general expression
\begin{equation}
\label{ }
\alpha(\omega) = \frac{1}{-m \omega^2 - i \omega  \tilde{\mu}(\omega) + K},
\end{equation}
with
\begin{equation}
\label{ }
{\rm Re}[ \tilde{\mu}(\omega)]=\frac{2e^2\om^2}{3c^3}|f(\bk_{0})|^{2} . 
\end{equation}
It follows in general, therefore, that the optical theorem is satisfied regardless of the specific choice for the form factor $f(\bk)$.

We conclude that, unlike the ALD theory, the approach to radiative reaction presented in Reference \cite{Ford91} results
in a polarizability that is consistent with all three basic physical requirements referred to in this paper, namely  causality, the crossing relation, and the optical theorem.  In addition, although the FO polarizability is mathematically different with respect to the ALD polarizability, one can easily check that the FO result is still consistent with
the familiar expression for the Rayleigh scattering cross section. \\

\noindent {\bf Acknowledgement}

The work of R. F. O'Connell was partially supported by the National Science Foundation under Grant No. ECCS-0757204.
F. Intravaia and R. Behunin acknowledge support from LANL's LDRD program.


\begin{thebibliography}{10}
\bibitem{note1} As discussed in Reference \cite{Ford91}, for instance, the results are quite insensitive to the specific
choice for $\Omega$.
\end{thebibliography}
\end{document}